# Diffusion-weighted MR spectroscopy: consensus, recommendations and resources from acquisition to modelling.

Clémence Ligneul*[1], Chloé Najac*[2], André Döring[3], Christian Beaulieu[4], Francesca Branzoli[5], William T Clarke[1], Cristina Cudalbu[6,7], Guglielmo Genovese[8], Saad Jbabdi[1], Ileana Jelescu[9,10], Dimitrios Karampinos[11], Roland Kreis[12,13], Henrik Lundell[14,15], Małgorzata Marjańska[8], Harald E. Möller[16], Jessie Mosso[17,6,7], Eloïse Mougel[18], Stefan Posse[19,20], Stefan Ruschke[11], Kadir Simsek[3,21], Filip Szczepankiewicz[22], Assaf Tal[23], Chantal Tax[3,24], Georg Oeltzschner[25,26], Marco Palombo[3,21], Itamar Ronen[27], Julien Valette[18]

**Affiliations:**
[1] Wellcome Centre for Integrative Neuroimaging, FMRIB, Nuffield Department of Clinical Neurosciences, University of Oxford; [2] C.J. Gorter MRI Center, Department of Radiology, Leiden University Medical Center, The Netherlands; [3] Cardiff University Brain Research Imaging Centre (CUBRIC), School of Psychology, Cardiff University, Cardiff, UK; [4] Departments of Biomedical Engineering and Radiology, University of Alberta, Edmonton, Alberta, Canada; [5] Paris Brain Institute - ICM, Sorbonne University, UMR S 1127, Inserm U 1127, CNRS UMR 7225, F-75013, Paris, France; [6] CIBM Center for Biomedical Imaging, Switzerland; [7] Animal Imaging and Technology, Ecole Polytechnique Fédérale de Lausanne, Lausanne, Switzerland; [8] Center for Magnetic Resonance Research, Department of Radiology, University of Minnesota, Minneapolis, Minnesota, USA; [9] Department of Radiology, Lausanne University Hospital, Lausanne, Switzerland; [10] Faculty of Biology and Medicine, University of Lausanne, Lausanne, Switzerland; [11] Department of Diagnostic and Interventional Radiology, Technical University of Munich, Munich, Germany; [12] MR Methodology, Department for Diagnostic and Interventional Neuroradiology, University of Bern, Switzerland; [13] Translational Imaging Center (TIC), Swiss Institute for Translational and Entrepreneurial Medicine, Bern, Switzerland; [14] Danish Research Centre for Magnetic Resonance, Centre for Functional and Diagnostic Imaging and Research, Copenhagen University Hospital - Amager anf Hvidovre, Denmark; [15] Department for Health Technology, Technical University of Denmark, Lyngby, Denmark; [16] NMR Methods & Development Group, Max Planck Institute for Human Cognitive and Brain Sciences, Leipzig, Germany; [17] LIFMET, EPFL, Lausanne, Switzerland; [18] Université Paris-Saclay, CEA,CNRS, MIRCen, Laboratoires des Maladies Neurodégénératives, Fontenay-aux-Roses, France; [19] Department of Neurology, University of New Mexico School of Medicine, Albuquerque, NM 87131; [20] Department of Physics and Astronomy, University of New Mexico School of Medicine, Albuquerque, NM 87131; [21] School of Computer Science and Informatics, Cardiff University, Cardiff, UK; [22] Medical Radiation Physics, Clinical Sciences Lund, Lund University, Lund, Sweden; [23] Department of Chemical and Biological Physics, The Weizmann Institute of Science, Rehovot, Israel; [24] University Medical Center Utrecht, Utrecht, The Netherlands; [25] Russell H. Morgan Department of Radiology and Radiological Science, The Johns Hopkins University School of Medicine, Baltimore, MD, United States; [26] F. M. Kirby Center for Functional Brain Imaging, Kennedy Krieger Institute, Baltimore, MD, United States; [27] Clinical Imaging Sciences Centre, Brighton and Sussex Medical School, Brighton, UK

**Corresponding authors:** clemence.ligneul@ndcn.ox.ac.uk, c.f.najac@lumc.nl

**Acknowledgements**: This paper was initiated by the Workshop "Best Practices & Tools for Diffusion MR Spectroscopy", held at the Lorentz Center (Leiden, Netherlands) on September 2021, and benefited from discussions among all (online and onsite) attendees (https://www.lorentzcenter.nl/best-practices-en-tools-for-diffusion-mr-spectroscopy.html). The workshop was also supported by NeurATRIS (A Translational Research Infrastructure for Biotherapies in Neurosciences, "Investissements d'Avenir", ANR-11-INBS-0011).

**Conflict of interest:** The authors declare no conflicts of interest.

**Submitted to Magnetic Resonance in Medicine**

# Graphical abstract

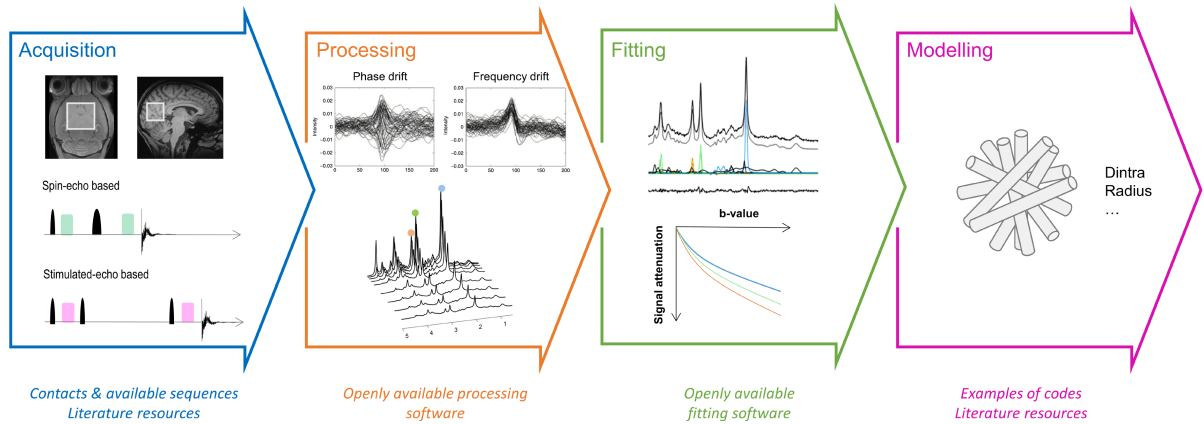

# Abstract

Brain cell structure and function reflect neurodevelopment, plasticity and ageing, and changes can help flag pathological processes such as neurodegeneration and neuroinflammation. Accurate and quantitative methods to non-invasively disentangle cellular structural features are needed and are a substantial focus of brain research. Diffusion-weighted MR spectroscopy (dMRS) gives access to diffusion properties of endogenous intracellular brain metabolites that are preferentially located inside specific brain cell populations. Despite its great potential, dMRS remains a challenging technique on all levels: from the data acquisition to the analysis, quantification, modelling and interpretation of results. These challenges were the motivation behind the organisation of the Lorentz Workshop on "Best Practices & Tools for Diffusion MR Spectroscopy" held in Leiden in September 2021. During the workshop, the dMRS community established a set of recommendations to execute robust dMRS studies. This paper provides a description of the steps needed for acquiring, processing, fitting and modelling dMRS data and provides links to useful resources.

# Abbreviations

| | |
|---|---|
| MRI | Magnetic Resonance Imaging |
| dMRI | Diffusion-weighted MRI |
| dMRS | Diffusion-weighted Magnetic Resonance Spectroscopy |
| MRS | Magnetic Resonance Spectroscopy |
| CNS | Central nervous system |
| tNAA | total *N*-acetylaspartate |
| NAA | *N*-acetylaspartate |
| NAAG | *N*-acetyl-aspartyl-glutamate |
| Glu | Glutamate |
| tCho | Choline compounds |
| PCho | Phosphorylcholine |
| GPC | Glycerophosphocholine |
| Ins | *Myo*-inositol |
| tCr | total Creatine |
| Cr | Creatine |
| PCr | Phosphocreatine |
| ADC | apparent diffusion coefficient |
| STE | STimulated Echo |
| TM | Mixing Time |
| $t_d$ | Diffusion time |
| SE | Spin Echo |
| TE | Echo Time |
| b-value | Diffusion- weighting (unit: ms/µm$^2$) |
| STEAM | STimulated Echo Acquisition Mode |
| sLASER | semi-LASER, LASER=Localization by Adiabatic SElective Refocusing |
| PRESS | Point RESolved Spectroscopy |
| RF | Radiofrequency |
| EC | Eddy currents |
| SAR | Specific Absorption Rate |
| DW | Diffusion-Weighted |
| CSF | Cerebrospinal fluid |
| FLAIR | Fluid-Attenuated Inversion Recovery |
| (B)PFG | (Bipolar) Pulsed Field Gradients |
| OG | Oscillation gradients |
| DDE | Double Diffusion Encoding |
| Tau | Taurine |
| NMR | Nuclear Magnetic Resonance |
| FA | Fractional Anisotropy |
| DTS | Diffusion Tensor Spectroscopy |
| $b_{max}$ | Maximum b-value used in an experiment |
| MM | Macromolecules |
| SNR | Signal-to-noise ratio |
| VOI | Voxel Of Interest |
| ECC | Eddy currents correction |
| $b_0$ | Minimum b-value required for spoiling |
| SVD | Singular Value Decomposition |
| FID | Free Induction Decay |
| MC | Metabolite-Cycling |
| TR | Repetition time |
| LCM | Linear-combination modelling |
| RMSE | Root-Mean-Square-Error |
| Gln | Glutamine |
| Glc | Glucose |
| ROC | Randomly-Oriented infinitely-long straight *Cylinders* |
| $B_0$ | Magnetic field strength |
| µFA | Microscopic Anisotropy |
| GM | Grey Matter |
| WM | White Matter |
| $D_{intra}$ | diffusivity inside the cylinders or sticks |
| $R$ | radius of the cylindrical structures comprising the ROC model |
| SGP | Short Gradient Pulse |
| ROS | Randomly Oriented Sticks |
| DTI | Diffusion Tensor Imaging |
| $\sigma$ | std of the Gaussian noise underpinning the normalised dMRS signal decay |
| Kapp | Apparent Kurtosis |
| CV | Coefficient of Variation |
| PGSE | Pulsed-Gradient Spin Echo |

## Introduction

The discovery that MRI can probe the diffusion process, and that the water in biological tissue was a ready-made, endogenous probe of tissue microstructure, ushered in an exciting era of innovation and development in diffusion-weighted MRI (dMRI). dMRI has numerous applications in neuroscience and clinical research. It also has several applications in clinical routine, including detecting stroke[1] and diagnosing prostate cancer[2]. With standard clinical MRI scanners, dMRI achieves spatial resolution on the order of millimetres thanks to the abundance of water in tissue, but by mapping diffusion measures dMRI can reveal tissue structure on a microscopic length-scale (microstructure). However, a challenge is the ubiquity of water in tissue and the heterogeneous nature of tissue microstructure, which leads to ambiguity in the interpretation of dMRI measurements. Almost concurrently with dMRI, dMRS was developed to exploit the compartmental specificity of MR-detectable molecules other than water for diffusion-based microstructural investigations.

Early metabolite-focused investigations used phosphometabolites, by implementing diffusion-weighted phosphorus magnetic resonance spectroscopy ($^{31}$P-dMRS) in the brain and muscle[3,4]. Subsequently $^1$H-dMRS measurements were made using endogenous molecules with discernible $^1$H resonances, which are thus detectable and quantifiable with $^1$H magnetic resonance spectroscopy (MRS) techniques. In the brain, for example, there are more than 10 detectable metabolites in the mM concentration range (**Figure 1A**). Most of them are predominantly intracellular, and therefore provide a probe of intracellular cytomorphology. Some metabolites have non-uniform distributions across cell-types. For example, the makeup of metabolites is different for neurons and astrocytes. Therefore, the relative concentrations of metabolites can be used to distinguish different cell-types (**Figure 1**). Initially measurements were done *in vitro*[5], and later *in vivo* in animal models[6,7] and in humans[8]. These early dMRS studies delivered microstructural information on healthy and diseased tissue, otherwise unattainable with dMRI[9]. Subsequent and significant advances in MR hardware, acquisition techniques, modelling strategies, and computational abilities, have allowed dMRS studies to provide unique insights on cellular morphology and physiology in health and disease. The history is eloquently summarised in several published reviews[10-14]. In parallel, a better understanding of the factors that affect measurement accuracy in *in vivo* dMRS experiments led to more robust and reproducible measurements of diffusion metrics. This enabled meaningful dMRS studies in clinical populations and in animal models (see **Figure 2**). The result is a steady increase in the number and variety of dMRS studies across the MR community, also covering applications in tissues beyond the central nervous system (CNS) (c.f. **Box 1**).

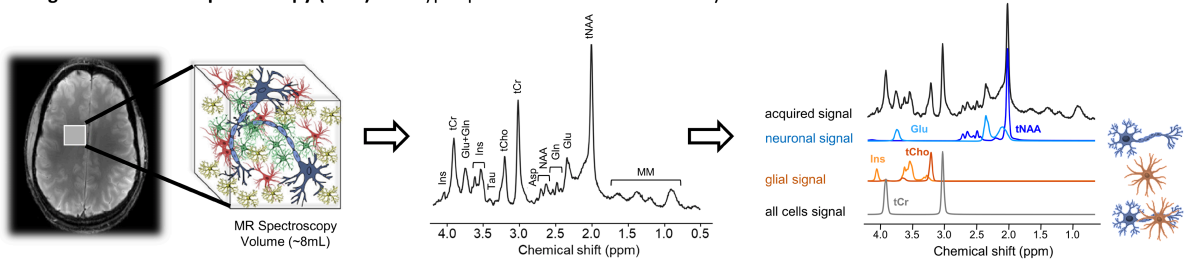

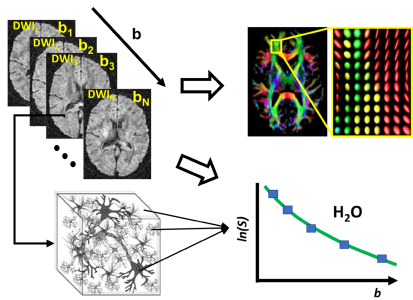 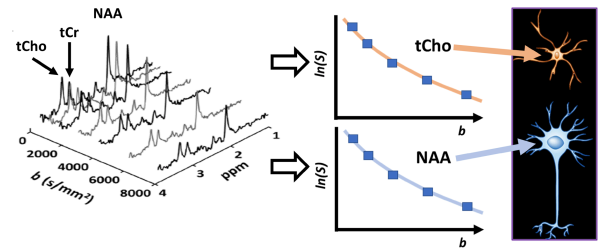

**Figure 1**: (A) Magnetic resonance spectroscopy (MRS) is the only neuroimaging tool providing *in vivo* non-invasive information on concentration of intracellular neuronal and glial metabolites. While total *N*-acetylaspartate (tNAA=NAA+NAAG) and glutamate (Glu) are found primarily in neurons, choline compounds (tCho=PCho+GPC) and *myo*-inositol (Ins) are mainly located in glial cells. Other metabolites, such as total creatine (tCr=Cr+PCr) are found in all cells. (Figure adapted from Palombo et al.[14]). (B) Diffusion-weighted imaging (dMRI) probes neural tissue microstructure via sensitization to the diffusion of water in tissue. Analysis of dMRI data provides microstructural information at high spatial resolution, but with no compartmental specificity, as water is distributed across cells and tissue compartments. (C) Diffusion-weighted MRS (dMRS) is based on sensitization of MRS to diffusion in a similar manner to dMRI. The diffusion properties of neuronal (e.g., NAA) and glial (e.g., tCho) metabolites reflect the specific microstructural environment of their host cell type.

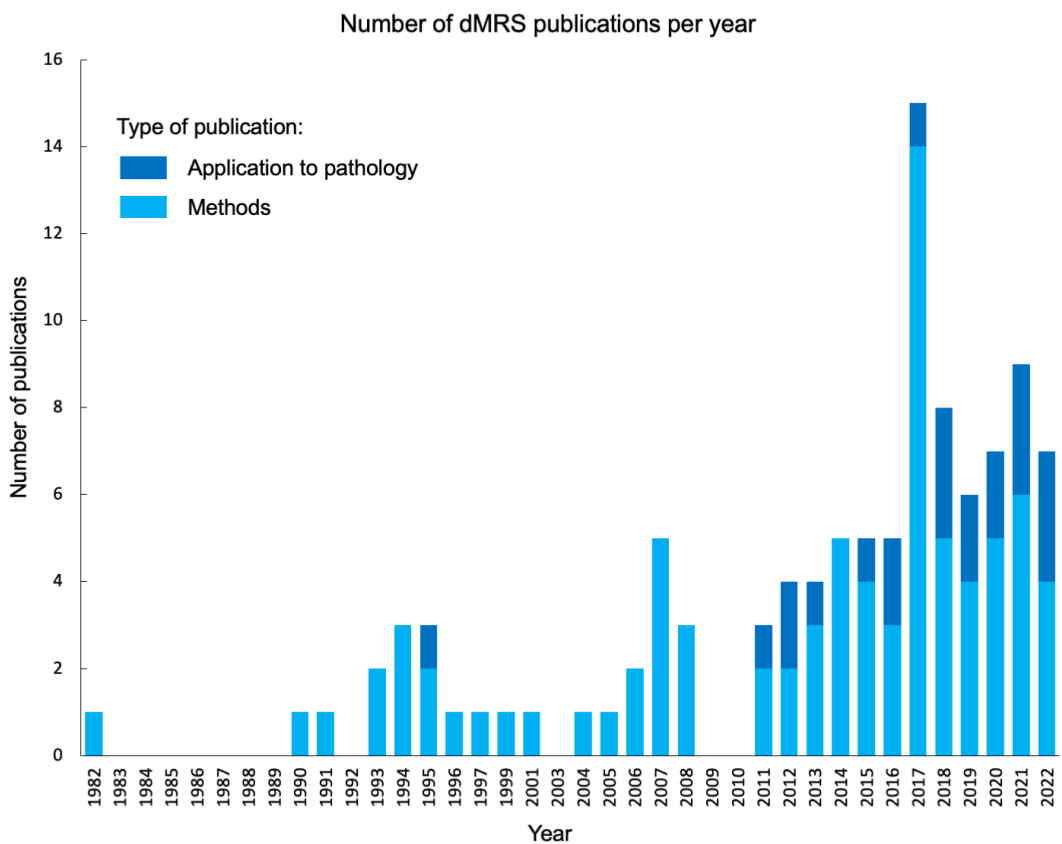

**Figure 2:** Number of dMRS publications per year. The dark blue tip corresponds to publications where dMRS is used to explore a brain disease (human) or a model of brain disease (animal). These publications were identified with a PubMed search initially yielding 196 results, curated to 114 results. Search criteria were: "diffusion weighted magnetic resonance spectroscopy" OR "dw-mrs" OR "diffusion-weighted mrs" OR "diffusion mrs" OR "diffusion weighted MR spectroscopy" OR "metabolites diffusion" OR "diffusion weighted NMR spectroscopy" OR "diffusion MR spectroscopy" OR "diffusion NMR spectroscopy" OR "diffusion magnetic resonance spectroscopy" OR "diffusion tensor spectroscopy". All chemistry papers (typically identifying protein structures) were manually removed. Yoshizaki et al., Biophys J, 1982; Van Zijl et al., PNAS, 1991; Merboldt et al., MRM, 1993; Van Gelderen et al., J. Mag. Res., 1994; Wick et al., Stroke, 1995; Pfeuffer et al., NMR Biomed, 1998; Pfeuffer et al., JCBFM, 2000; Shemesh et al., Nature Comm, 2014; Branzoli et al., NMR Biomed, 2014; &Lundell et al., NeuroImage, 2021 were manually added. These papers are referenced on the MRShub dMRS forum[a] and in dMRS references and educationals[b].

Despite the great promise of a specific, non-invasive tool for measuring *in vivo* cytomorphology, the use of dMRS in preclinical and clinical investigations remains limited. Most dMRS studies are performed by a small number of sites that are involved in developing dMRS methodology, or by groups in direct collaboration with experts from these sites. These interesting studies are typically small, proof-of-concept studies that show the potential of dMRS. However, the sparse adoption of dMRS indicates that methods are not yet simple and robust enough to garner traction from independent and non-expert users.

---

[a]https://forum.mrshub.org/t/educationals-getting-started/946/3;
[b]https://docs.google.com/spreadsheets/d/1SS1mz3fDQfW-OjdhPAjb5iSGPjuVjm5MyzVyLz9EngE/edit#gid=0

There are several challenges for a widespread adoption of dMRS. Foremost is the absence of robust, vendor-provided dMRS pulse sequences on commercial (human and animal) scanners, and second is a similar absence of readily available pipelines for processing and analysing dMRS data. In addition, three other challenges may deter researchers from engaging in dMRS studies. First, is the perception among researchers and clinicians that dMRS studies are complex and difficult to perform. Second, dMRS results are inconsistent across sites, originating from sensitivity to differences in acquisition, processing, and analysis methods. This gives the impression that dMRS is not reliable enough, for example, to provide biomarkers of disease. Finally, the long acquisition time of dMRS experiments suggests that dMRS is incompatible with investigations in clinical populations.

These challenges were the motivation behind the organisation of the [Lorentz Workshop on "Best Practices & Tools for Diffusion MR Spectroscopy"](#)[c] held in Leiden in September 2021. The community of dMRS developers, together with that of MRS quantification experts has established a set of common goals to make all components of the dMRS pipeline, from acquisition via processing and analysis to modelling, readily available and openly accessible.

This paper provides step-by-step recommendations that were identified during the workshop for planning and executing a robust dMRS study. Throughout this paper, we will assume that the reader has knowledge of basic MRS concepts and is interested in incorporating dMRS in their research protocols. The focus is on the brain, but most recommendations from Sections 1 to 3 hold for non-CNS applications (c.f. **Box 1**). In the first section, we provide an overview of dMRS pulse sequences used for data acquisition, and we clarify under which conditions each should be used. We provide guidelines for choosing diffusion-weighting schemes and optimal diffusion-weighting parameters, depending on the desired outcome of the experiment. In the second section, we lay out the processing steps that are required to maximise data quality by minimising controllable inaccuracy and uncertainty affecting dMRS metrics. The third section is dedicated to practices related to spectral fitting of dMRS data for optimal retrieval of dMRS metrics such as metabolite apparent diffusion coefficients (ADC). The fourth and final section discusses key aspects of the dMRS signal in relation to brain microstructure and proposes the randomly oriented infinite cylinders model as an acceptable minimal biophysical model, within a given applicability framework. Finally, some examples of future directions for dMRS are listed.

---

[c]https://www.lorentzcenter.nl/best-practices-en-tools-for-diffusion-mr-spectroscopy.html

**Box 1: dMRS outside the brain**

Although most dMRS applications are focused on the CNS, methods have also been developed in the body for various applications[15]:

- To measure water diffusion properties, which are not confounded by fat[16,17]
- To measure fat properties within adipose tissue[18]. Best acquisitions schemes are described in[19-21]. Specific vibration-compensation schemes that might be required are described in[22]
- To measure diffusion of lipid molecules such as intra- and extramyocellular lipids and less abundant metabolites within skeletal muscle[23,24]
- To separate overlapping peaks such as lactate and lipids[25-27]

# 1. dMRS acquisitions

In this section, we will cover the acquisition of dMRS data, spanning from the choice of the sequence to the experimental setup itself.

## 1.1. Choosing the right pulse sequence

The choice of the dMRS sequence is governed by the research question (e.g., what parameters are of interest) and the locally available sequences and hardware (e.g., what are the limitations). dMRS beginners can find contacts for available sequences on the dMRS section on MRShub[d].

Stimulated Echo *versus* Spin Echo:

There are two broad categories of DW-sequences: those based on a stimulated echo (STE), where the diffusion-weighting gradient moment builds up during the mixing time TM (being essentially the diffusion time, $t_d$); and those based on a spin echo (SE), where diffusion-weighting gradient moment builds up entirely within the echo time (TE).

The advantage of STE-based sequences is that very large b-values can be achieved, while maintaining short TE. In addition, STE-based sequences enable long $t_d$ (particularly important when probing slowly diffusing metabolites and large diffusion restriction barriers). However, half of the signal is lost when using STE. The advantage of SE is essentially that it retains full signal. However, in practice, reaching long $t_d$ or very large b-values requires long TE, which will in turn result in significant signal loss due to $T_2$ relaxation (and due to *J*-modulation for metabolites such as glutamate (Glu) or *myo*-inositol (Ins)).

Recommendations about the choice of sequence are summarised in Table 1. Animal and human scanners have different hardware capabilities, hence different pulse sequences are presented here.

Pulse sequences for human scanners:

STE-based diffusion measurements on human scanners are preferably performed using STEAM (STimulated Echo Acquisition Mode)[8,28] (**Figure 3A**). Because long TM enables high b-values while keeping diffusion-weighting gradient strengths relatively low, e.g., not larger than spoiler or slice-selection gradients, cross-terms might be significant and must be taken into account (see **Box 2**).

For SE-based volume selection, the sLASER (semi-LASER) sequence[29] (where LASER stands for "Localization by Adiabatic SElective Refocusing"[30]) is now generally preferred over PRESS

---

[d]https://forum.mrshub.org/t/sequences-and-acquisitions-strategies-for-dw-mrs/934

("Point RESolved Spectroscopy"), as it is robust to $B_1^+$ inhomogeneities and offers large radiofrequency (RF) pulse bandwidth to minimise chemical shift displacement error. Diffusion-weighting gradients can be inserted in a "double bipolar" fashion[10], as shown in **Figure 3C**, thus maximising diffusion-weighting, and compensating for eddy currents (EC).

Table 1: Overview of recommended sequences for human and animal scanners, depending on the research question. (B)PFG = (bipolar) pulsed field gradients, OG = oscillating gradients, DDE = double diffusion encoding; tNAA = *N*-acetylaspartate + *N*-acetylaspartylglutamate, tCr = creatine + phosphocreatine, tCho = phosphocholine + glycerophosphocholine, Glu = glutamate, Ins = *myo*-inositol, Tau = taurine.

| Interested in | Stimulated echo or spin echo? | Human scanner | Animal scanner |
|---|---|---|---|
| **Mostly singlets (tNAA, tCho, tCr)** | Spin echo | BPFG-sLASER | SE-LASER |
| ***J*-coupled metabolites (Glu, Ins, Tau)** | Stimulated echo | PFG-STEAM | STE-LASER |
| **Very high b-values** | Stimulated echo | PFG-STEAM | STE-LASER |
| **Long diffusion times** | Stimulated echo | PFG-STEAM | STE-LASER |
| **Advanced encoding: OG, DDE** | Spin echo | OG/DDE-sLASER | OG/DDE-LASER |

Sequences for small animal scanners:
dMRS sequences tailored for humans can also be used in small animal scanners. However, one can take advantage of rodent scanners' specificities to alleviate some limitations of human scanner sequences. In particular, stronger gradients and less conservative Specific Absorption Rate (SAR) limits enable the use of shorter TE and additional RF pulses. This allows a block-based sequence design where diffusion-weighting and localization are independent, resulting in the absence of cross-terms (**Box 2**) between diffusion and localization gradients. In practice, the localization is achieved using a full LASER module, while the diffusion block is built around non-selective STE[31] or SE[32,33] (**Figure 3D**).

About water:
Water suppression schemes are the same as for conventional MRS acquisition. However, when metabolite signals are too weak in individual transients for estimating scan-to-scan frequency, phase, and amplitude alterations (see **Section 2: Processing**), preserving the water signal with a metabolite cycling approach is recommended to observe both water and metabolites. This is already available for diffusion-weighted-STEAM (DW-STEAM) and DW-sLASER[34-36]. Another option is to de-optimize water suppression so that some clean residual signal is still visible. Because water signal arising from cerebrospinal fluid (CSF) may contribute substantially at low b-values (<~1 ms/µm²) and may not represent metabolite signal distortions properly, it can be suppressed with a water-selective fluid-attenuated inversion recovery (FLAIR) module at the start of the sequence[35,37] while preserving water signal from parenchymal tissues.

### 1.2. Experimental setup and challenges

Optimising data quality

Methods for acquisition of optimal MRS data have been described in recent consensus papers[38,39] and strategies to improve the quality of spectra with motion compensation have been published[40]. Nevertheless, in the context of dMRS, some parameters that can influence the diffusion measurements, and must be managed very carefully, are summarised in Table 2 and detailed below when necessary.

*Pilot study on phantom:* It is strongly recommended to test a newly implemented dMRS sequence on a phantom before acquiring data *in vivo*. For a standard set of low b-values, it is possible to use a simple agarose phantom (water alone in a phantom is more prone to mesoscopic flows and not ideal, particularly for longer $t_d$). However, metabolites' diffusion is about 4-5 times slower than that of water, and a water-based phantom cannot be used to assess the quality of a sequence at high b-values (> 3-4 ms/µm²). Ethylene glycol is a cheap liquid that has a diffusion coefficient at room temperature of about 0.1 µm²/ms and a simple NMR spectrum. Its use is recommended at high b-values[31]. On human scanners, it is also possible to use the standard "Braino" phantom[41] for b < 1.6 ms/µm², or a National Institute for Standards and Technology (NIST) phantom[42] to access a wider range of diffusivities. A few basic measurements to evaluate in such freely diffusing liquids are:
- a monoexponential decay of the signal as a function of the b-value (when changing gradient strength), corroborating the expected diffusion coefficient (at room temperature, agarose phantom or water-based phantom: ~2 µm²/ms, ethylene glycol phantom: ~0.1 µm²/ms),
- a constant diffusion coefficient when the $t_d$ is changed,
- isotropy of the above measurements.

If these observations are not matched, the sequence needs to be revised (often pointing towards a miscalculation of the b-value and cross terms, see **Box 2**).

Experimental devices/tools

*Animal setup (c.f. Table 2):* A stable anaesthesia level should be maintained to avoid potential bias[43]. Animal holders and fixation are excellent tools to minimise motion during the acquisition.

Sequence parameters and acquisition management

*Handling signal accumulation via repeated transients (c.f. Table 2):* Care should be taken to perform phase cycling instead of having the same phase on all transients (which might be the case by default for some systems when storing individual transients, e.g., when using "number of repetitions" in Bruker's ParaVision®).

There is no consensus as to whether the number of transients per b-value should be increased with b-value. There are, however, arguments for doing so, when practically feasible. First, some fitting softwares, such as LCModel[44] (see **Section 2**), will provide more consistent signal quantification if averaged spectra across b-values have a comparable signal-to-noise ratio (SNR), but more robust fitting tools such as FiTAID with 2D prior knowledge may not require the same consistency. Second, more transients might be corrupted by motion artefacts at high b-values, which will then be discarded during processing. So one may want to acquire more transients to start with. On the other hand, acquiring enough transients at high b-values to match low b-value SNR might be too time-consuming, especially in human applications. Optimised acquisition schemes can be further tested with simulations depending on the application, on the fitted model, on the available time for the acquisition, etc. Some authors have already proposed optimised b-value schemes for an accurate estimation of NAA diffusion coefficient in the corpus callosum, using coefficient of variation as a criterion, yet without considering a varying number of transients per b-value[45].

*Volume-of-interest (VOI) size (c.f. Table 2):* It should be carefully adjusted to yield just the minimally acceptable SNR in the region of interest, typically to allow for the main metabolite peaks to be visible on individual transients at the highest b-value to allow for efficient frequency and phase correction (see **Section 2**). However, too large a voxel may result in artifactual signal attenuation due to bulk rotation (since, for a given rotation angle, the absolute displacement at the extremity of the voxel is larger for a larger voxel, hence resulting in larger phase variation).

*Cardiac triggering:* While small translational motion can be corrected for by adjusting individual frequency/phase shifts during processing (see **Section 2**) or by using prospective motion correction methods[40], the effect of physiological motion (cardiac / CSF pulsation) might result in amplitude drops, particularly in humans at high b-values[46,47] and for VOIs located close to the ventricles. The use of cardiac-triggering with an optimal trigger delay in combination with retrospective processing correction (see **Section 2.2.**) allows to minimise bias. FLAIR to null water signal from CSF when using metabolite cycling is not feasible in combination with triggering. Similarly, cardiac triggering places limits on the length of the water suppression module, and may also cause TR to be different from transient to transient and increase acquisition time. In body applications, both cardiac and respiratory triggering might be necessary.

*Water suppression*: Diffusion-weighting is rather helpful regarding water suppression. Since water diffuses faster than metabolites, diffusion gradients strongly attenuate the water signal. However, when the TM is very long (typically to study $t_d$ > 500 ms), the water suppression might degrade. Although the water signal should be completely crushed before excitation, if a small amount of water was not crushed properly, this residual signal will relax during the TM. The crushers scheme has to be carefully optimised, but it is sometimes challenging. In practice, adding a water suppression pulse during the TM can help crush the residual signal when using a long TM.

Additional acquisitions required for processing

*Macromolecules*: For a TE shorter than ~80-100 ms, contributions of macromolecules (MM) cannot be neglected[48], especially at high b-values, where contributions of MM become more prominent relative to metabolites due to slow MM diffusion (Figure 4 in[14]). It is recommended to acquire an experimental MM spectrum for subsequent unbiased quantification of metabolites (see **Section 3**). This is achieved using metabolite-nulling by inversion-recovery (or double inversion-recovery)[49]. Suppression of residual metabolite signal can be further improved by performing the acquisition at high b-values (typically 10 ms/μm²)[50] or as described in[48] but always using the specific dMRS sequence for acquisitions. MM signals can also be used as an endogenous sensor for motion[51], see **Section 2**.

*Eddy current correction* (ECC) may be required, in particular on human scanners when high gradient strength is used. ECC relies on measuring the temporal phase of the water signal[52]. Hence, unless metabolite cycling is used, non-water-suppressed spectra must be acquired as reference for all diffusion conditions requiring ECC. In that case, it is important that the water signal dominates these reference scans. Note that this is not always possible at high b-values due to faster water diffusion, unless the water-suppressed spectrum is subtracted from the non-water-suppressed spectrum[53]. An increased number of transients also ensures a high water SNR:

a noisy reference scan used for ECC will introduce artefacts during processing. In applications where water peak-based ECC is not possible, suppression of EC effects may be feasible using a bipolar diffusion-encoding scheme under the assumption of a linear system response where EC effects cancel out for opposite readout polarities[25,54].

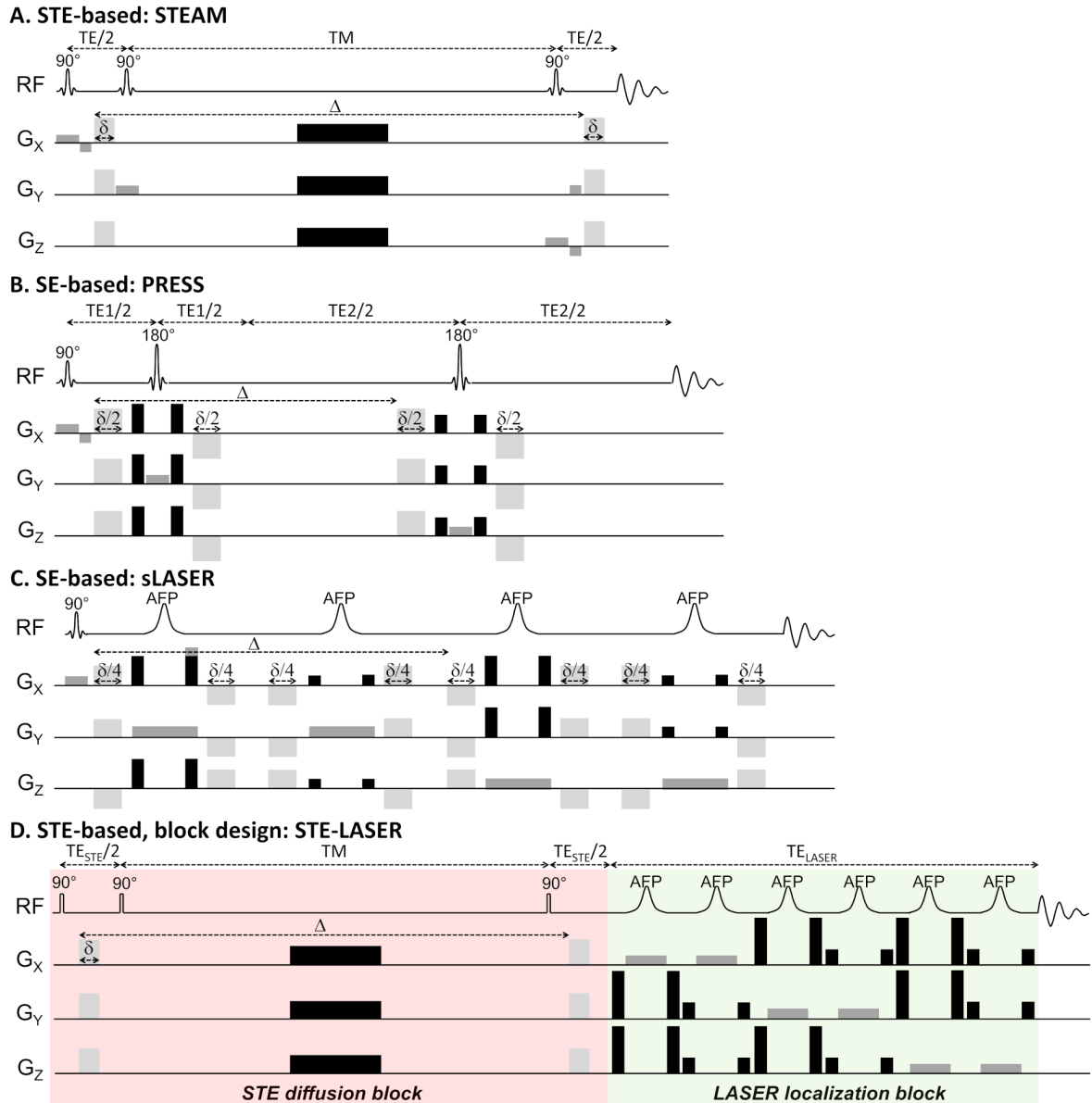

**Figure 3:** Chronograms of typical dMRS pulse sequences. (A) STE-based STEAM, using a pulsed field gradient scheme[8]; (B) SE-based PRESS, using a bipolar pulsed field gradient scheme[55]; (C) SE-based sLASER, using a double bipolar pulsed field gradient scheme[47]; (D) STE-based block-design STE-LASER, using a pulsed field gradient scheme[31]. The STE block (in pink) can be replaced by a SE block. While the first three sequences can be used in humans, the last one is difficult to achieve in humans due to SAR and TE. Figure adapted from Palombo et al.[14].

Table 2: Checklist for a successful dMRS acquisition. (*) contain more details in the main text. This table only describes what is specific to dMRS in comparison with MRS. The prior is a successful MRS acquisition.

| **Experimental devices/tools** | |
|---|---|
| **Temperature** | *In animals* (*): monitor and maintain stable body temperature |
| **Motion** | - Store individual transients for correction during processing<br>- *In humans*: use cardiac-triggering (+ respiratory-triggering for applications outside the CNS)<br>- *In animals*: use appropriate animal holder |
| **Vibration (gradient-induced)** | *Suggestions for STE-acquisition*: possible to add an additional set of gradients preceding the excitation, and matching the diffusion encoding gradients and diffusion time |
| **Sequence parameters and acquisition management** | |
| **Minimal b-value** | Ensure minimal gradient strength required for efficient spoiling of spurious echoes |
| **b range** | - *ADC measurement*: max b-value ~3-5 ms/µm$^2$<br>- *Kurtosis*: max b-value ~8-10 ms/µm$^2$<br>- *Fully dispersed neurites model:* max b-value ~20 ms/µm$^2$ ($t_d$ value is also important, cf. Section 4)<br>- *Bi-exponential analysis:* max b-value ~25 ms/µm$^2$ |
| **Number of directions** | - For rotation invariant powder averaging (cf. Section 4)<br>   ○ 3 directions minimum in a grey matter voxel with low fractional anisotropy (FA)<br>   ○ 12 directions in a white matter voxel with high FA<br>- For *diffusion tensor spectroscopy* (DTS)<br>   ○ 6 directions minimum<br>   ○ 3 directions to compute the trace only |
| **Transients** (*) | Store individual transients for phase and frequency correction in processing |
| **Voxel size** (*) | Large enough to detect one singlet signal (typically tNAA) at $b_{max}$ in single shot (for processing), unless water signal can be used in a metabolite-cycling experiment |
| **Interleaved acquisitions** (*) | Mitigates biases due to frequency, shimming, sensitivity drifts, and metabolic variations |
| **Phase cycling** | Ensure proper phase cycling across b-values and diffusion encoding directions, in particular if single acquisitions are discarded based on data quality |

| Additional acquisitions required for processing | |
|---|---|
| **Macromolecules (MM)** (*) | Acquire MM using inversion recovery + high b-value to eliminate residual metabolite signal |
| **Eddy current correction** (*) | Acquire water signal with identical gradient scheme as metabolite scans or when not possible data with opposite gradient polarity |

**Box 2: "cross-terms"**

Diffusion gradients are the main contributor, but other gradients in the sequence also contribute to the diffusion-weighting. As described in[56], the calculation of the b-value follows:

$$b = -\gamma^2 \int_0^{T_{seq}} \left[ \int_0^{t''} (G_{diff}(t') + G_{others}(t'))dt' \right]^2 dt''$$

when $G_{diff}$ (diffusion gradients) and $G_{others}$ (other gradients, typically localisation gradients and/or spoilers) are intertwined, like in the STEAM-based and sLASER-based dMRS sequences, a cross term $\int G_{diff} \int G_{others}$ appears. To estimate the true b-value, this term has to be calculated and taken into account, as it grows with the diffusion gradient strength (unlike $\int G_{others} \int G_{others}$ that is a constant contribution at all b-values).

The calculation of the true b-value can be incorporated directly into the sequence code or calculated a posteriori based on the sequence chronogram. Examples of codes are openly available [on MRS hub][e] to numerically compute the b-value based on the chronogram. It is possible to check in a phantom whether cross-terms should be accounted for.

Another popular approach to compensate for cross-terms consists in acquiring data with diffusion gradients of opposite polarities[57,58]. The geometric mean of these signals is free from cross-terms, although strictly speaking this is only valid in the limit of mono-exponential attenuation and under the assumption of homogenous diffusion within the voxel. Another advantage of this method is that it also compensates for any static background gradients that cannot be foreseen. Alternatively, the diffusion encoding gradient waveforms can be designed to minimise the error due to cross-terms[59].

In general, the contribution of cross-terms should be kept minimal, as they can lead to a poorly defined diffusion time and affect the modelling of microstructural features.

---

[e] https://forum.mrshub.org/t/processing-for-dw-mrs/935

## 2. Processing steps specific to dMRS data

dMRS and general MRS data processing share most steps. As an excellent consensus paper already exists on MRS processing[60], the objective of this section is to emphasise where additional care is required for dMRS. A robust and accurate processing is particularly important for dMRS. Indeed the estimation of diffusion measures like ADC relies on the measurement of the attenuation of the signal across multiple acquisitions (each with different diffusion-weighting conditions) and is thus prone to error propagation. For example, a 5% underestimation of the signal of a given metabolite at b=$b_0$+3 ms/µm$^2$ will lead to a 15% overestimation of its ADC in a typical measurement setup (ADC ~ 0.12 µm$^2$/ms; 30% expected signal attenuation between $b_0$ and $b_0$+3 ms/µm$^2$). At lower b-values (e.g., $b_0$+1 ms/µm$^2$), the same error on metabolite quantification can lead to a 42% overestimation of the ADC. Although precision is better at higher b-values, the accuracy of ADC estimation is corrupted by kurtosis from b=3-5 ms/µm$^2$ (see **Section 4.2**). Moreover, diffusion gradients make diffusion-weighted spectra even more sensitive to the subject's motion and scanner instabilities. Processing can help mitigate the effect of small bulk or physiological motion artefacts, which result in frequency/phase drifts (translational motion) and amplitude fluctuation (rotation, compressive motion)[13]. An overview of the current toolboxes available for processing (and fitting, see **Section 3**) dMRS data is provided in Table 3.

The dMRS community tends to use custom-made Matlab routines, passed on and incrementally modified by users. From a survey carried out during the 2021 Lorentz workshop, most pipelines contained the recommended processing steps detailed in this section. Although some of these pipelines are shared along with publications on a git platform, they are not necessarily easy to find for newcomers. In Table 3, we provide a summary of the available toolboxes/packages that can be used for dMRS data processing. Since the functions to be used might vary with package updates, more details about how to use these packages for dMRS processing can be found in the dMRS section on MRShub[e].

---

[e]https://forum.mrshub.org/t/processing-for-dw-mrs/935

Table 3: Available packages referenced on MRShub to process and/or fit MRS data. Packages highlighted in dark grey are routinely used by dMRS users. An orange tick mark means that the package is freely available, but relies on a licensed language. A package is considered open source when a shared repository exists. *jMRUI 7.0 is to be released soon, with FitAID as a plugin **for Osprey basis set generation comes from the online tool MRSCloud. The updated version of this table can be found on [MRShub dMRS forum](https://forum.mrshub.org/t/processing-for-dw-mrs/935/2)[f].

| | Processing | | | | Fitting | | | | | | |
|---|---|---|---|---|---|---|---|---|---|---|---|
| | Coils Combination | Phase & frequency correction | ECC | Water residual removal | Basis set generation | Signal fitting Sequential | 2D | Open source | Free | Language | Reference |
| ABfit | ✗ | ✗ | ✗ | ✗ | ✗ | ✓ | ✗ | ✓ | ✓ | R | Wilson, 2021[61] |
| AQSES | ✗ | ✓ | ✓ | ✓ | ✗ | ✓ | ✗ | ✓ | ✓ | Java | Poullet at al., 2007[62] |
| FID-A | ✓ | ✓ | ✓ | ✓ | ✓ | ✗ | ✗ | ✓ | ✓(orange) | Matlab | Simpson et al., 2017[63] |
| FitAid | ✗ | ✗ | ✗ | ✗ | ✗ | ✓ | ✓ | ✗ | ✓ | Java | Chong et al., 2011[64] |
| FSL-MRS | ✓ | ✓ | ✓ | ✓ | ✓ | ✓ | ✓ | ✓ | ✓ | Python | Clarke et al., 2021[65] |
| jMRUI 6 * | ✗ | ✓ | ✓ | ✓ | ✓ | ✓ | ✗ | ✗ | ✓ | Java | Stefan et al., 2009[66] |
| LCModel | ✗ | ✗ | ✓ | ✗ | ✗ | ✓ | ✗ | ✓ | ✓ | Fortran | Provencher, 1993[44] |
| Osprey | ✓ | ✓ | ✓ | ✓ | ✓ | ✓** | ✗ | ✓ | ✓ | Matlab | Oeltzschner et al., 2020[67] |
| ProFit-1D & SpectroS | ✓ | ✓ | ✓ | ✓ | ✓ | ✓ | ✗ | ✓ | ✓ | Matlab | Borbath et al., 2021[68] |
| spant | ✓ | ✓ | ✓ | ✓ | ✓ | ✓ | ✗ | ✓ | ✓ | R | Wilson, 2021[69] |
| Tarquin | ✓ | ✓ | ✓ | ✓ | ✓ | ✓ | ✗ | ✓ | ✓ | C++ | Wilson et al., 2011[70] |
| Vespa | ✓ | ✓ | ✓ | ✓ | ✓ | ✓ | ✗ | ✓ | ✓ | Python | Soher et al., 2011[71] |

[f]https://forum.mrshub.org/t/processing-for-dw-mrs/935/2

## 2.1. dMRS specificities for processing steps shared with conventional MRS

For the following steps, the reader can refer directly to[60]:

- RF coil combination (2.2.1 in[60])
    - The amplitude, phase, and noise terms necessary for coil combination can generally be determined from the $b_0$ unsuppressed water data for dMRS.
- Phase and frequency drifts correction (2.1.3 in[60])
    - The effect of phase and frequency correction is particularly important at high diffusion weighting when the phase shift between scans is larger (**Figure 4**).
- Outlier removal (2.1.2, Table 2, Figure 4 in[60])
    - The outlier removal must be carried out at a given set of b-value and direction. If more than 30% of the transients are affected, it is advised to discard the dataset. See **Section 2.2.**
    - After the individual transients are averaged and sorted, it is still possible that the resulting spectrum is corrupted by motion. See **Section 2.3.** for further motion detection.
- ECC (see **Section 1**)
    - When water data at high b-value are too noisy for ECC, an alternative (during processing) is to decouple the water signal from the noise with a singular value decomposition (SVD) and use the synthetic free induction decay (FID) for the ECC.

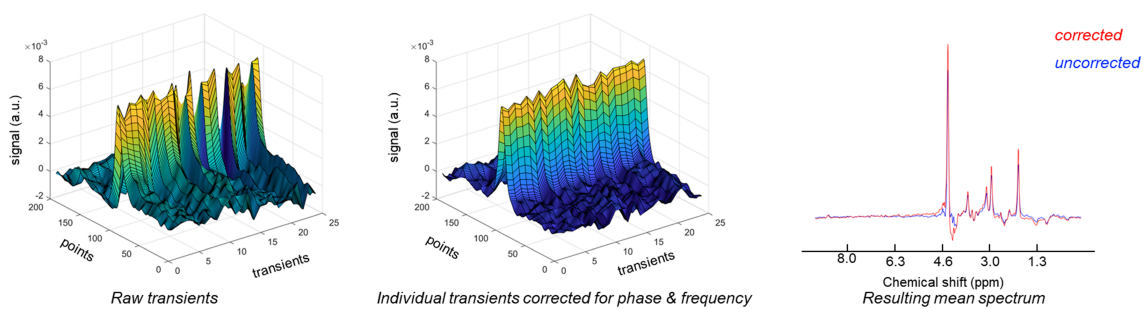

**A. Example of phase and frequency correction for b ~ 0.6 ms/µm²**

*Raw transients* — *Individual transients corrected for phase & frequency* — *Resulting mean spectrum*

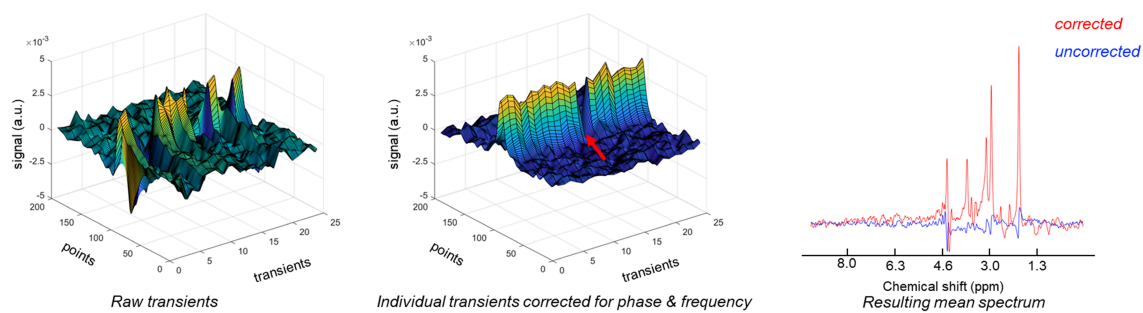

**B. Example of phase and frequency correction for b ~ 4 ms/µm², with one outlier** (*red arrow*)

*Raw transients* — *Individual transients corrected for phase & frequency* — *Resulting mean spectrum*

**Figure 4:** Examples illustrating the importance of phase and frequency correction at (A) b=0.6 ms/µm² and (B) at b=4 ms/µm²

## 2.2. Specific considerations for a metabolite cycling approach

With a metabolite-cycling (MC) approach (see **Section 1**), the non-suppressed water signal can be beneficial for **compensating motion-related signal** loss[35].

Since this technique does not use water suppression, it is susceptible to sideband artefacts[72] that can affect spectral quality if any hardware component is malfunctioning and fluctuating, e.g., with the power source. Therefore spectra should be carefully inspected for artificial resonance peaks, which appear at the base modulation frequency or its higher harmonics (e.g., multiples of 50 Hz). However, due to its 10,000-fold higher SNR, water is a particularly useful internal reference not only to improve coil-channel averaging, phase and eddy current correction, but also to compensate for signal loss due to motion[35]. This motion compensation is based on the assumption that a certain quantile of transients at a given b-value remains unaffected by motion (e.g., 25% quantile of transients) and thus can serve as an internal reference for rescaling motion-induced decay/dephasing. When using cardiac triggering (see **Section 1**), the effective repetition time (TR) may vary across transients and thus induce $T_1$-related amplitude fluctuations between b-values if not taken into account. It is then suggested to apply a correction factor $(1-\exp(-T_{Reff}/T_1))$ prior to defining the reference level[73].

Note that motion compensation may fail when all transients are compromised by motion (e.g., if only a small number of transients is acquired, if all transients are similarly distorted due to table vibrations, or if at extremely high b-values water has decayed to the level of metabolite amplitudes).

## 2.3. Specific considerations for short TE dMRS: macromolecules

Acquisition of a MM spectrum is briefly described in **Section 1** and more detailed information can be found in[48]. Some MM spectra are openly available (MRShub)[g]. For dMRS, some specific details about MM should be considered: at very high b-values, MM account for half of the signal (Figure 4 in[14]). *In vivo* MM should be acquired with the same sequence as the dMRS signals using a single or double inversion recovery module. The MM signal that is used for fitting needs a very low noise level and experts in the field use two different methods: acquiring a very high number of MM spectra until the noise level is barely detectable, or fitting the average MM spectrum[74], and including the resulting fit (noise-free by default, but requires confidence in the fit) in the basis set.

MM can also be used as an internal probe for residual motion effects for averaged spectra (per DW condition). MM have a very low ADC (~ 0.005 - 0.01 µm$^2$/ms), hence their signal is virtually unaffected by diffusion-weighting. To evaluate residual motion effects, one can visually inspect stack plots of spectra (before fitting) for gross signal drop, or assess changes in MM signal by integrating its peak at 0.9ppm. It can be done prior to fitting by inspecting visually stack plots of spectra for gross signal drop, or by integrating the MM peak at 0.9 ppm[51]. If the MM fit is reliable, it can also be done after spectral fitting, by measuring the MM ADC. The MM decay is monoexponential up to very high b-values[31].

## 3. Spectral fitting for estimating metabolite diffusion parameters

Quantitative information is extracted from an acquired MR spectrum by fitting a physics-informed model to it. The expert consensus recommends the use of linear-combination modelling (LCM), usually based on non-linear least-squares optimization, for most *in vivo* MRSapplications[60] and is also recommended for dMRS. The objective of this section is to discuss and emphasise aspects of spectral fitting of particular importance to dMRS, and suitable methods to incorporate diffusion modelling into the spectral fitting process. Along with a careful processing (c.f. **Section 2**), a well-designed fitting pipeline can mitigate critical issues routinely encountered with dMRS such as motion dephasing, lineshape distortions, eddy currents, and frequency drift[75].

---

[g]https://mrshub.org/datasets_mm/

## 3.1. Basis sets and macromolecular signals

Diffusion weighting changes metabolite-specific signal amplitudes, not J-evolution. Thus, metabolite basis sets are unaffected by diffusion encoding and can be simulated as for conventional MRS[60]. Note that for spin echo sequences longer TEs required to achieve strong diffusion weighting will exhibit greater fit uncertainty due to low SNR, particularly for overlapping metabolite signals[76]. To ease the fitting of dMRS data, predefined basis sets are increasingly made publicly available and will be collected and curated by the MRS community (MRS Hub, basis-sets[h]). When these predefined basis sets are used one has to take care to select those matching the dMRS sequence and TE.

For TE<80-100ms it is advised to include an experimentally measured MM pattern in the basis set (c.f. **Section 1**)[48]. If experimentally measured MM are not available, the MM signals should be modelled with parameterized individual MM components ('MM09', 'MM12', etc.) that are included in many contemporary fitting algorithms. To reduce the degrees of freedom and prevent variability of final ADC estimation, the parameterized MM should be derived at the highest b-value and kept constant in shape for the other diffusion-weightings.

## 3.2. Sequential and simultaneous spectral fitting

DW spectra span an interrelated 2D space defined by the spectral (time/frequency) and diffusion (dephasing) dimension. Both dimensions have to be addressed: (i) the spectral dimension is decomposed into contributions from individual metabolite patterns (basis sets) by LCM[60]; (ii) the signal decay along the diffusion dimension is fitted to the metabolite amplitude estimates, either using an abstract diffusion signal representation (e.g., mono-/biexponential, kurtosis, DTS) (see **Section 3.2.1.**) or a signal model directly related to tissue microstructure (e.g., cylinders, spheres, multi-compartment)[77] (see **Section 4**). To achieve accurate fitting along the diffusion dimension, correct b-values should be calculated, including slice-selective and crushing gradients, and cross-terms (c.f. **Box 2**).

With recently available software, a dMRS dataset can be fitted either in a *sequential* (1D) or *simultaneous* (2D) mode. In sequential mode, fitting along the diffusion dimension is performed in a separate post-hoc optimization after *sequential (traditional) spectral fitting*, whereas *simultaneous fitting* explicitly incorporates the diffusion model into the spectral fitting process. *Simultaneous fitting* is not a subject of consensus or recommendation yet but is a very promising approach, hence the underlying concepts are introduced in this part.

*3.2.1. Common signal representations*

---

[h]https://mrshub.org/datasets_basissets/

There are different diffusion signal representations possible, depending on the acquisition parameters (b-values sampled, $b_{max}$, number of directions). It has become established practice to use a monoexponential representation for b≤5 ms/µm², a kurtosis or biexponential for b≤10 ms/µm² and a biexponential for b>10 ms/µm²[31]. The root-mean-square-error (RMSE) of the residuals of the biexponential fit for a specific metabolite can serve as a predictor for its standard deviation from a Gaussian noise model. It is a useful measure to determine whether or not a microstructural model is applicable (C.f. **Section 4**, Table 5). However, it is important to note that the fast and slow diffusion components cannot be interpreted as individual cellular compartments. If at least 6 directions are acquired, it is also possible to compute the diffusion tensor from each monoexponential decay[78]. For the MM, a monoexponential representation is usually sufficient though at very high b-values this assumption may start to break down[31]. These representations describe diffusion using only a few parameters (2 for monoexponential, 3 for kurtosis, and 4 for biexponential), which prevents overfitting (though introducing constraints on metabolite diffusion, which can affect tissue modelling).

*3.2.2. Sequential (1D) fitting*
In *sequential fitting* the spectrum arising from each separate diffusion encoding (b-value, direction, $t_d$, or encoding scheme) is fitted independently. Most fitting tools include baseline estimation (e.g., adding a smooth spline or polynomial), and the baseline flexibility is a major contributor to uncertainty and variance in the fitting process[79,80]. Therefore, care should be taken to prevent the baseline from introducing additional variability and systematic differences into the estimates of metabolic diffusion properties at different b-values. It is advisable to not over parameterize the baseline and keep it similar between b-values with the same number of spline knots or low-degree polynomials (e.g., to include residual water artefacts). If possible, relative metabolite frequency shifts in individual spectra should be constrained to a minimum to prevent inconsistent models at different b-values and limit the degree of freedom of the overall fitting process.

Fit results for metabolites with similar spectral patterns are difficult to reliably separate and it is common practice to indicate the total combined estimated area for correlated groups of metabolites (e.g., tNAA = NAA + NAAG, tCr = Cr + PCr, tCho = PCho + GPC). Separate estimates of strongly overlapping signals will have lower precision. The use of such sums in diffusion analysis implies that the diffusion properties of the constituents are comparable, but that is not at all a given, either because the molecules are of substantial difference in size and molecular weight (e.g., PCho and GPC, and Cr and PCr) or cellular location (e.g., Gln (glutamine) and Glu). If the diffusion properties are expected to be similar, it may be beneficial to use soft constraints to regularise the relative concentration ratios of such connected metabolites to a predefined value to obtain more consistent results and prevent overfitting. If diffusion properties differ, this approach will fail.

After sequential spectral fitting, the diffusion-weighted individual metabolite amplitude estimates or their sums (one per diffusion encoding) can be fitted using a signal representation or a tissue model[77].

### 3.2.3. Simultaneous (2D) fitting

In *simultaneous fitting,* all spectra from all diffusion encodings are fitted in a single step[75]. Spectral parameters are adapted for each b-value, direction, $t_d$ or encoding scheme under the constraints of separate priors on the selected diffusion model for each metabolite or for the MM. For example, in an experiment with increasing b-value, amplitude decays for each metabolite could be constrained to a (multi-)exponential decay function along the diffusion dimension. As in general, simultaneous fitting constraints can be applied to width, phase, or relative frequency parameters to be consistent for the whole data set – though care should be taken not to over-constrain the model given that lineshapes may alter based on EC (possibly not fully compensated and thus dependent on diffusion gradient strengths). Simultaneous fitting allows to introduce assumptions on equal diffusion behaviour for multiple metabolites to stabilise the fits at each b-value (e.g., equal ADCs for metabolites like *scyllo*-inositol and Ins; or monoexponential decay for the MM; or even more complex models for particular metabolites). Moreover, the use of concentration ratio constraints for overlapping metabolites with differing diffusion properties is feasible and may promise to help segregation of metabolite patterns with differing diffusion properties (though without sufficient SNR spurious effects from overfitting could still result when allowing parameters to adapt). Simultaneous fitting likely also aids baseline estimation (by constraining its parametrization to a specific decay model) and lineshape parametrization (which is difficult at high b-values due to low SNR)[73].

The biexponential representation is good to stabilise *simultaneous fitting* for b>10 ms/µm² since it represents the signal well, is easy to implement, and has a minimum number of free parameters. It is useful when fitting diffusion properties of J-modulated or low concentration metabolites like Glu, Ins or glucose (Glc). Indeed, a recent study on a set of synthetic dMRS spectra found a more reliable estimate of diffusion and tissue properties when simultaneous fitting is applied[81]. This is supported by Tal et al.[82] showing that *simultaneous 2D fitting* improved precision by at least 20% compared to 1D fitting.

## 3.3 Comparability of results between fitting algorithms and methods

It should be noted that fitting of *in vivo* MRS data represents a complex, ill-posed optimization problem, and there are many available algorithms to solve it. Recent studies have revealed that fitting results can vary substantially not just between fitting algorithms[80,83,84], but also within any given algorithm[83,84], depending on highly sensitive fit options and settings. Preliminary evidence from the 2021 dMRS workshop suggests that dMRS is no exception to this and given the strong impact of error propagation on the estimated ADC, achieving optimal spectral fitting performance and continued comparison across methods is critical. As the field of dMRS develops, it will be important to benchmark and validate emerging fitting methods, ideally against synthetic data with known ground truth.

# 4. Modelling - Linking the diffusion properties to tissue microstructure

In this section, we will focus on the consensus and recommendations regarding the biophysical modelling of purely intracellular metabolites to infer histologically meaningful descriptions of brain microstructure from dMRS. We highlight three points of consensus that set out the currently accepted minimal model of metabolites' diffusion in brain tissue. We then provide guidelines on how to adapt this minimal model to typical experiments.

## 4.1 The minimal model of metabolite diffusion in brain tissue: Randomly-Oriented infinitely-long straight Cylinders (ROC)

The ROC model is a simple analytical model of diffusion in infinitely long (i.e. diffusion is only restricted in transverse directions) and randomly oriented cylinders (i.e. reflecting mesoscopic isotropy). This section describes why it is a reasonable biophysical model to extract microstructural information from metabolite diffusion, and conditions that need to be fulfilled for it to be applicable.

*4.1.1 Intracellular brain metabolites exhibit non-Gaussian diffusion*

At high b-values and for typical diffusion lengths of 5-15 µm, the dMRS signal from intracellular brain metabolites exhibits non-mono-exponential decay (**Figure 5A**). This provides a hint about the length scale of structures hindering metabolite diffusion (a few µm). Mono-exponential behaviour breaks for b-values > 3-5 ms/µm$^2$. This has first been observed with early measurements of NAA diffusion in brain cells at various $t_d$ and to high b-values (up to b=30 ms/µm$^2$)[85]. Other intracellular metabolites exhibit similar non-mono-exponential decays, as reported in subsequent studies[31,86]. Multi-exponential (usually bi-exponential) signal representations sometimes used to characterise the signal decays at high b-values do not inform about cytoarchitecture. Biophysical models are thus built to extract meaningful cytoarchitectural parameters.

*4.1.2 Correlation between relaxation and diffusion properties is negligible for intracellular brain metabolites*

Correlation between diffusion and relaxation properties may arise from the presence of different compartments, each with its own relaxation times and diffusion properties (e.g., cytosolic versus mitochondrial compartments). The studies summarised in Table 4 suggest that for 35≲TE≲70 ms, TM≲250 ms and $t_d$ ≲250 ms, minimal to no correlation between relaxation and diffusion properties exist for intracellular brain metabolites (**Figure 5A**). Therefore, the diffusion measurements performed within this regime carry information about the cellular microarchitecture, unbiased by the inherent relaxation processes. This also suggests that experiments performed within this experimental regime should yield comparable results across studies.

Table 4: Summary of studies reporting potential dependence between the TE used and the measured diffusion properties.

| Publications | $B_0$ field & Sample | TE range (ms) | $t_d$ (ms) | Conclusions |
|---|---|---|---|---|
| Assay & Cohen, 1998[87] | 11.7 T *Ex vivo* rat brain | 70-200 | 35 & 95 | NAA bi-exponential fit depends on TE |
| Assaf & Cohen, 1999[85] | 11.7 T Excised bovine optic nerve | 70-550 | 125 | NAA signal decay depends on TE |
| Branzoli et al., 2014[88] | 7 T *In vivo* human brain | 40-160 | 44-246 | No TE dependence for TE < 90 ms (except tCr at $t_d$=246 ms). |
| Ligneul et al., 2017[31] | 11.7 T *In vivo* mouse brain | 33-73 | 20-253 | No TE dependence for ADC or ADCs of fast and slow components (from bi-exponential fit) |
| Mougel et al., 2022[89] | 11.7 T *In vivo* mouse brain | 50-110 | 20 | No TE dependence for ADC and Kurtosis (except very small ADC variation for tCr) |

*4.1.3 Purely intracellular brain metabolites exhibit signatures of microscopic anisotropy.*
The non-Gaussian nature of metabolite diffusion can have different origins. Experimental evidence supports the hypothesis that one of the major sources is the microscopic anisotropy (µFA) of the brain microstructure. Microscopic anisotropy represents the level of anisotropy at the microscopic scale (i.e., in the range of the diffusion length probed, 5-15 µm) and not at the (macroscopic) voxel size scale (known in dMR as fractional anisotropy, FA). For instance, if metabolites were mostly located in round cell bodies, the measured µFA and FA would be close to 0. On the other hand, if they were mostly located in thin and elongated fibres, randomly oriented within the spectroscopic voxel, then measured µFA would be close to 1, but FA would still be close to 0.

Several studies using single diffusion encoding acquisitions and high b-values have demonstrated that the diffusion of intracellular metabolites exhibits signatures of microscopic anisotropy, both in rodent grey matter (GM)[90,91] and in human white matter (WM)[92]. More recently, from powder-averaged data acquired at high b-values, it has been measured that the µFA of NAA nears 1 in human WM[93]. A few works exploring metabolite ADC over a wide range of $t_d$ also suggest that metabolites are primarily located in elongated fibres[94-96].

DDE acquisitions are challenging but provide more direct measures of µFA. General considerations for DDE in dMRI and dMRS have been covered in recent reviews[97,98]. Pioneering DDE studies reported high µFA for all intracellular metabolites[32,99], but slightly higher for NAA (a neuronal metabolite) than for Ins (a glial metabolite). More recent work reports similar results both in the rodent and the human brain[100,101].

The diffusion of intracellular metabolites is anisotropic at the microscopic scale in both GM and WM and should therefore represent a key component for any biophysical model of metabolite diffusion (**Figure 5B**).

**A. Intracellular brain metabolites exhibit non-Gaussian diffusion and correlation with relaxation properties is weak.***

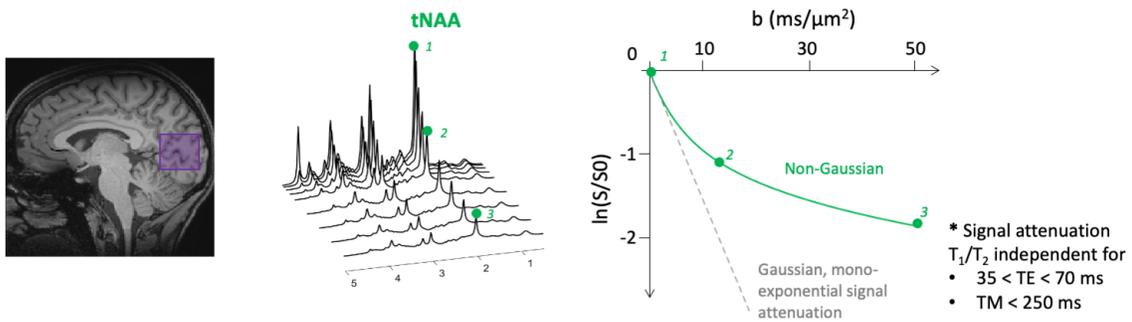

* Signal attenuation $T_1/T_2$ independent for
- $35 < TE < 70$ ms
- $TM < 250$ ms

**B. Micro-anisotropy of intracellular brain metabolites is high and their diffusion properties match with diffusion in elongated fibres.**

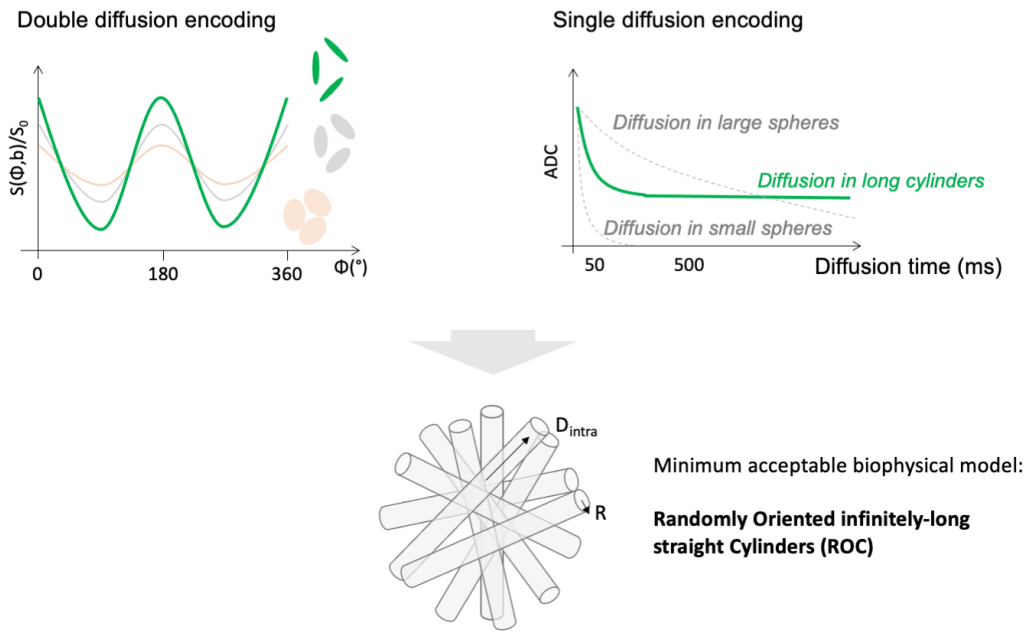

Minimum acceptable biophysical model:

**Randomly Oriented infinitely-long straight Cylinders (ROC)**

**Figure 5.** (A) Intracellular metabolites exhibit non-Gaussian diffusion, and correlation with relaxation properties is weak. The grey curve (right panel) represents the logarithm of the normalised signal decay in the case of Gaussian diffusion (e.g., free diffusion). The green curve (right panel) represents the logarithm of the normalised signal decay, coming from tNAA. The data were simulated for the Lorentz workshop pregame, and are available on [github][i]. Data always originates from large voxels, as exemplified in the left panel. (B) Intracellular brain metabolites µFA is high and their diffusion properties match with diffusion in elongated fibres. (left, green curve) Intracellular metabolites exhibit a strong modulation with the angle ϕ between diffusion directions in DDE-MRS experiment, reflecting a high µFA. (right, green curve) Their ADC from very short to very long $t_d$ is well represented by diffusion in long cylinders (right, grey curves) in contrast to diffusion in spheres. (Bottom) The ROC model is characterised by the diffusivity $D_{intra}$ along the cylinder and the radius $R$. If $t_d$ is too long and/or $b_{max}$ is too low, the acquisition does not support the estimation of the radius $R$, and the model can be simplified to randomly oriented sticks, with $D_{intra}$ as the only parameter.

---

[i]https://github.com/dwmrshub/pregame-workshop-2021

*4.1.4 ROC model*

The previous sections set out the reasons why the ROC model is a reasonable "minimal" biophysical model for brain metabolite diffusion. It effectively mimics the geometry of cellular processes (neuronal and glial), which is where metabolites are primarily located (only being minimally located in cell bodies).

- *Analytical expression* : the signal attenuation $K$ from a given cylinder with orientation **n** by a diffusion gradient pair with orientation **g** can be written as a function of the acquisition parameters (**g**, $\Delta$, $\delta$) and the model parameters $D_{intra}$ (intracellular diffusivity) and $R$ (cylinder radius), either using the Short Gradient Pulse (SGP) approximation[102] or the van Gelderen's formula in the case of longer pulses[9]. The orientation distribution of the cylinders can be then factored out by averaging the signal over a reasonably large number of diffusion directions, the so-called directional average or powder-average, $S$:

$$S(g, \delta, \Delta; D_{intra}, R) = \frac{\int_0^\pi sin(\theta)\, K(g, \theta, \delta, \Delta;\, D_{intra}, R) d\theta}{\int_0^\pi sin(\theta) d\theta}$$

  leaving $D_{intra}$ and $R$ to be estimated. Here $\theta$ is the angle between the direction of the cylinder's main axis **n** and the direction of **g**.

- *Open-source implementations of the ROC model:* widely used in the dMRI community, several toolboxes contain implementations of this model (e.g., MISST[j], DMIPY[k]). A list of these packages and examples of simple scripts can be found on MRS hub[l]. We also list educational resources to help get a feeling of the contributions of different parameters (like DIVE[m]).

*4.1.5 Conditions of applicability*

- *TE regime:* as described in 4.1.2, for data acquired with 35 ≲ TE ≲ 70 ms at 7 T ≤ $B_0$ ≤ 11.7 T, it is possible to ignore $T_2$ effects. Due to $T_2$ dependence to magnetic field, for $B_0$ ≤ 3 T the upper bound on TE is expected to be higher, for $B_0$ ≥ 11.7 T the lower bound on TE is expected to be higher.

- *Number of directions*: the number of directions needed to achieve a rotation invariant powder averaging increases with the term b·$D_{intra}$[103] and with macroscopic anisotropy in the voxel. In mostly isotropic voxels (such as in the rodent GM), as few as 3 directions might be sufficient [103]. In a very anisotropic voxel (such as in high FA human WM), with ideal SNR conditions, a total of 12 directions should provide a rotational invariance up to very high b-values. However, in practice, other noise sources might dominate when fewer than 12 directions are used, even in isotropic voxels[93].

---

[j]http://mig.cs.ucl.ac.uk/index.php?n=Tutorial.MISST; [k]https://github.com/AthenaEPI/dmipy;
[l]https://forum.mrshub.org/t/modeling-dw-mrs-data/943;
[m]https://git.fmrib.ox.ac.uk/fsl/DIVE/-/blob/master/README.md

- *Limitations and considerations on $t_d$*: for SE-based pulse sequences (see **Section 1**), the minimum $t_d$ and TE are often constrained by the gradient duration and must be kept constant unless $t_d$-dependence is explored. $t_d$-dependence goes beyond the scope of this paper and requires different modelling. To gain higher sensitivity to fibre radius *R*, shorter $t_d$ are preferred.

As general guidelines for the ROC analysis, we recommend designing the acquisition to reach a maximum b-value of ~20 ms/µm², keeping TE and $t_d$ constant and as short as possible.

For typical $t_d$ ~ 40-80 ms, the effect of branching (i.e. metabolites exchanging between cell ramifications) is negligible and exchange between soma and cellular fibres might be negligible[95,104]. However, cytoarchitectural features with a characteristic length scale comparable to a diffusion length along fibres (5-10 µm) might impact $D_{intra}$ estimated by the ROC model. These features could be[105]:

- *spines,* occurring at ~1-2 spines/µm,
- *undulations,* occurring over a wavelength ≳10-16 µm,
- *beading*/fibre *diameter variations* occurring over a length scale of 5-8 µm.

The impact of restriction due to *somas* of radius ≳12 µm might also be non-negligible[104].

For $t_d$ ≳ 100 ms, *branching* affects the diffusion properties[95,104] and sensitivity to the radius extracted from the ROC model is reduced, because the diffusion length relative to the radius is too big ($4t_d D_{intra} >> R^2$). Likewise, in cases of low $b_{max}$ (b-value<<$t_d/R^2$, e.g., b-value<<10 ms/µm² for fibre *R*=2 µm and $t_d$=40 ms) or low SNR, the signal attenuation cannot be used to extract the fibre radius from a ROC model (see **Section 4.2**). A model of randomly oriented sticks (ROS) should be considered instead (i.e., cylinders with zero radius):

$$K(b, \theta; D_{intra}) = exp[-bD_{intra}\cos(\theta)^2]$$

The powder-average signal can then be expressed as:

$$S(b, D_{intra}) = \sqrt{\frac{\pi}{4bD_{intra}}} \text{erf}(\sqrt{bD_{intra}})$$

where erf denotes the error function. If the sticks' orientations are uniformly distributed on the sphere (i.e., FA=0) the ADC along any arbitrary direction derived from low b-values is then $D_{intra}/3$.

- *Modified ROC model for very anisotropic voxels (e.g., WM):* The ROC model is based on the assumption that within a large spectroscopic voxel the cellular processes (i.e., neurite and glial processes) are randomly oriented. This condition is usually met for GM voxels or when using enough diffusion directions to yield a reliable powder average. However, it may not be met when using few diffusion directions in voxels located primarily within WM tracts, where the preferential orientation of the cell processes may significantly impact the measured diffusion signal. In this case, the ROC model should be modified to account for the specific orientation distribution of fibres within the spectroscopic voxel.

The simplest way to do so is to acquire dMRI data suitable for diffusion tensor imaging (DTI) analysis together with the dMRS data and estimate the distribution of principal fibres directions from the DTI analysis (as explained in the Supporting Information of Reference[91]). The fraction of WM fibres within the voxel, $f_{WM}$, can be estimated by counting the fraction of pixels in DTI images within the spectroscopic voxel and that have high FA (e.g., ≥0.4); and $c_L$≥0.4; $c_P$≤0.2 and $c_S$≤0.35[106]. $c_L$, $c_P$ and $c_S$ are the coefficients of linearity; planarity and sphericity, respectively. Note that this assumes maximal fibre anisotropy in WM (i.e., a worst-case scenario). The total signal attenuation for the modified ROC model then comprises two contributions: one coming from metabolites' diffusion in the GM (pure ROC), and the second coming from metabolites' diffusion in the WM modelled as diffusion within fibres distributed according to $P(\theta_i)$:

$$S(g, \delta, \Delta; D_{intra}, R)$$
$$= (1 - f_{WM}) \frac{\int_0^\pi \sin(\theta) \, K(g, \theta, \delta, \Delta; D_{intra}, R) d\theta}{\int_0^\pi \sin(\theta) d\theta}$$
$$+ f_{WM} \frac{\sum_{i=1}^N P(\theta_i) K(g, \delta, \Delta; D_{intra}, R, \theta_i)}{\sum_{i=1}^N P(\theta_i)}$$

where the distribution $P(\theta_i)$ is the distribution of angles $\theta_i$ between the first eigenvector of the diffusion tensor and the diffusion-sensitising gradients and is experimentally determined (hence the discrete summatory rather than a continuous integral). As already mentioned, if sensitivity to $R$ is poor, then the signal attenuation K should be modelled as diffusion within sticks.

## 4.2 Considerations of noise and maximum b-value

The accuracy and precision of the ADC, apparent Kurtosis (Kapp) and ROC model parameter estimates depend on $\sigma$, the standard deviation of the Gaussian noise underpinning the dMRS signal decay normalised to the signal at b=0 ms/µm$^2$. The $\sigma$ can be estimated for measurements, for example, by computing the standard deviation of the residuals from fitting the bi-exponential representation to normalised diffusion-weighted signals (Supplementary Figure 1A), or from denoising techniques that also provide an estimate of the noise level[107].

In Table 5, we show the simulated impact of noise on the relative bias and coefficient of variation (CV) of the different parameters. The ADC can be estimated accurately and precisely (i.e., both relative bias and CV median and interquartile range < 10%) with $b_{max}$ < 5 ms/µm$^2$ and $\sigma \leq 0.04$. In contrast, Kapp can be estimated accurately and precisely with $b_{max}$ < 10 ms/µm$^2$ only for low $\sigma \leq$ 0.01. Regarding the ROC model with $b_{max} \leq 25$ ms/µm$^2$, $D_{intra}$ can be estimated accurately and precisely for $\sigma \leq 0.04$, while $R$ can only be estimated accurately and precisely for very low $\sigma <<$ 0.01. If such a noise level cannot be achieved, we recommend using the ROS.

The general recommendation is therefore to keep $\sigma \leq 0.04$, to accurately estimate ADC (with $b_{max}$ < 5 ms/µm$^2$) and $D_{intra}$ (with $b_{max} \leq 25$ ms/µm$^2$); $\sigma \leq 0.01$ when trying to estimate Kapp (with $b_{max}$ < 10 ms/µm$^2$) and $\sigma <<$ 0.01 when trying to estimate $R$ (with $b_{max} \leq 25$ ms/µm$^2$). Choices made from acquisitions to fitting (Section 1-3) will impact $\sigma$ and should therefore be carefully setup.

Table 5. Simulated impact of Gaussian noise (for different standard deviation $\sigma$) on the relative bias and coefficient of variation (CV) of ADC, Kapp and ROC model parameters $D_{intra}$ and $R$. For each model parameter median [1st; 3rd] quartile values are reported. **Data simulation**: analytical simulations were performed using the ROC model assuming an acquisition that can be run on both clinical and preclinical scanners: Pulsed-Gradient Spin Echo (PGSE) acquisition with $\Delta/\delta$ = 60/15 ms; b = 0, 1.5, 3, 5, 8.5, 15, 25 ms/µm$^2$. The simulations were run for a range of realistic combinations of $D_{intra}$ (= 0.2, 0.25, 0.30, 0.35, 0.40, 0.45, 0.50 µm$^2$/ms) and $R$ (= 0.25, 0.50, 0.75, 1.00, 1.25, 1.50, 2.00 µm) values, with the addition of Gaussian noise corresponding to different $\sigma$ levels. **Impact of Gaussian noise on parameter estimates**: to estimate the fit accuracy at different $b_{max}$, a Monte Carlo approach (250 draws, Supplementary Figure 1B) was used to extract the median and the first and third quartiles of each parameter estimate. The fits were a mono-exponential decay for ADC, a kurtosis representation for ADC and Kapp, and a ROC model for $D_{intra}$ and $R$. Note that at each new draw, new noisy signal is created by adding Gaussian noise with the standard deviation estimated from the fitting residuals of the previous draw. Orange shadowed entries in the table highlight conditions where accuracy and precision was considered too low, i.e., relative bias and CV median and interquartile range >10%.

**Relative Bias: (Estimated - Ground Truth)/Ground Truth (%)**

| Representation or Model | $b_{max}$ | Parameter | $\sigma = 0.01$ | $\sigma = 0.02$ | $\sigma = 0.04$ | $\sigma = 0.1$ |
|---|---|---|---|---|---|---|
| Mono-exponential | 5 | ADC | -1 [-4; 1] | 3 [-2; 7] | 3 [-9; 14] | -9 [-27; 34] |
| | 10 | Not Applicable | - | - | - | - |
| | 25 | Not Applicable | - | - | - | - |
| Kurtosis | 5 | ADC | -2 [-7; 4] | -2 [-12; 6] | -4 [-19; 7] | 6 [-27; 26] |
| | 5 | Kapp | 3 [-78; 21] | -33 [-77; 22] | -32 [-87; 25] | -42 [-79; 12] |
| | 10 | ADC | -1 [-4; 3] | 0 [-5; 8] | 0 [-7; 10] | -1 [-16; 27] |
| | 10 | Kapp | -2 [-9; 1] | 3 [-16; 16] | -25 [-38; 12] | -40 [-78; 15] |
| | 25 | Not Applicable | - | - | - | - |
| | | | - | - | - | - |
| ROC | 5 | $D_{intra}$ | -7 [-22; -3] | -18 [-41; -4] | -31 [-45; -6] | -30 [-45; -10] |
| | 5 | R | 22 [-39; 193] | 113 [-17; 298] | 143 [-36; 333] | 178 [0; 391] |

| | 10 | $D_{intra}$ | -3 [-8; 1] | -7 [-16; 1] | -15 [-35; -5] | -29 [-52; -6] |
| | | R | 2 [-61; 98] | 44 [-34; 163] | 87 [-27; 232] | 217 [28; 486] |
| | 25 | $D_{intra}$ | -1 [-4; 0] | -2 [-7; 3] | -5 [-15; 2] | -18 [-34; 1] |
| | | R | 1 [-40; 39] | -3 [-46; 72] | 1 [-48; 136] | 90 [-7; 241] |

CV: std[Estimated]/mean[Estimated] (%)

| Representation or Model | $b_{max}$ | Parameter | $\sigma = 0.01$ | $\sigma = 0.02$ | $\sigma = 0.04$ | $\sigma = 0.1$ |
|---|---|---|---|---|---|---|
| Mono-exponential | 5 | ADC | 4 [2; 5] | 5 [3; 7] | 11 [8; 16] | 26 [18; 38] |
| | 10 | Not Applicable | - | - | - | - |
| | 25 | Not Applicable | - | - | - | - |
| Kurtosis | 5 | ADC | 4 [3; 6] | 8 [5; 11] | 13 [9; 19] | 40 [23; 61] |
| | | Kapp | 39 [13; 149] | 74 [32; 191] | 99 [34; 290] | 119 [51; 224] |
| | 10 | ADC | 4 [3; 5] | 7 [5; 9] | 11 [9; 14] | 21 [17; 31] |
| | | Kapp | 10 [6; 16] | 14 [10; 34] | 47 [23; 73] | 97 [35; 185] |
| | 25 | Not Applicable | - | - | - | - |
| | | | - | - | - | - |
| ROC | 5 | $D_{intra}$ | 11 [6; 26] | 26 [17; 35] | 34 [25; 40] | 47 [37; 58] |
| | | R | 99 [52; 156] | 77 [41; 176] | 99 [53; 242] | 100 [44; 198] |
| | 10 | $D_{intra}$ | 5 [3; 7] | 12 [8; 17] | 21 [14; 33] | 41 [31; 50] |
| | | R | 76 [42; 162] | 88 [55; 170] | 95 [71; 179] | 72 [39; 166] |
| | 25 | $D_{intra}$ | 3 [2; 4] | 6 [4; 7] | 12 [8; 15] | 35 [23; 42] |
| | | R | 49 [18; 121] | 73 [39; 130] | 85 [43; 166] | 84 [58; 136] |

## Future directions

dMRS forms a lively research field that has taken off in the last 10 years. Although dMRS faces challenges regarding SNR, it has a great potential for probing, non-invasively, cell-type-specific microstructure, and could benefit directly from the latest developments in MRS and in dMRI. A few challenges awaiting solutions in acquisition, processing, fitting, and modelling are listed below. Solving these would unlock much of the potential of dMRS, both as a research tool and for clinical use. These developments are interdependent, and will benefit from each other:

- Faster scanning is crucial for clinical applications. A few approaches will make this possible:
  - Developing and validating simultaneous fitting approaches as proposed in [75] and integrated in FitAID[64] and FSL-MRS[65],
  - Taking advantage of the recent development of multiparametric acquisitions and use it to concurrently measure relaxation and diffusion for instance or better sample the q-$t_d$ space[108],
  - Further developing and validating robust denoising approaches[107],
  - Improving acquisitions methods (e.g. rapid diffusion tensor acquisition[109]).
- Improving the spatial resolution and coverage with diffusion MRSI[110,111].
- Developing cross-modalities approaches (e.g. dMRI and dMRS with a joint modelling of water and metabolites diffusion).
- Interpreting the data with a metabolic or microstructural perspective (e.g. continuing the development of modelling to extract microstructural parameters, incorporation into simultaneous fitting, and accounting for metabolic complexity etc.)

# Supplementary information

Supplementary Table 1: detail of author contributions (* equally contributed)

| Name | Contribution | | | | |
|---|---|---|---|---|---|
| | Workshop Organisation | Paper section leading | Paper writing | Paper substantial commenting / reviewing | Paper reviewing / endorsing |
| Clémence Ligneul* | x | x | x | x | |
| Chloé Najac* | x | | x | x | |
| André Döring | | x | x | x | |
| Christian Beaulieu | | | | | x |
| Francesca Branzoli | x | | x | | |
| William Clarke | | | | x | |
| Cristina Cudalbu | | | | x | |
| Guglielmo Genovese | | | | x | |
| Saad Jbabdi | | | | x | |
| Ileana Jelescu | | | x | x | |
| Dimitrios Karampinos | | | | | x |
| Roland Kreis | | | x | x | |
| Henrik Lundell | | | x | x | |
| Malgorzata Marjanska | | | | x | |
| Harald E. Möller | | | | x | |
| Jessie Mosso | | | x | | x |
| Eloïse Mougel | | | x | | x |
| Stefan Posse | | | | | x |
| Stefan Ruschke | | | x | | x |
| Kadir Simsek | | | x | | |
| Filip Szczepankiewicz | | | x | x | |
| Assaf Tal | | | | | x |
| Chantal Tax | | | | | x |
| Georg Oeltzschner | | | x | x | |
| Marco Palombo | x | x | x | x | |
| Itamar Ronen | x | x | x | | |
| Julien Valette | x | x | x | x | |

**Supplementary Figure 1**: Schematic to clarify the difference between (A) the level of noise σ of the normalised dMRS signal decay and (B) how Monte-Carlo simulations help estimating the model accuracy.

### A. Estimation of dMRS normalised signal decay noise σ

Example: using a bi-exponential fit for estimating **σ**, the level of noise in the data.

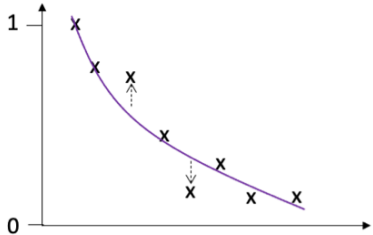

### B. Estimation of model accuracy for noise σ with Monte-Carlo simulations

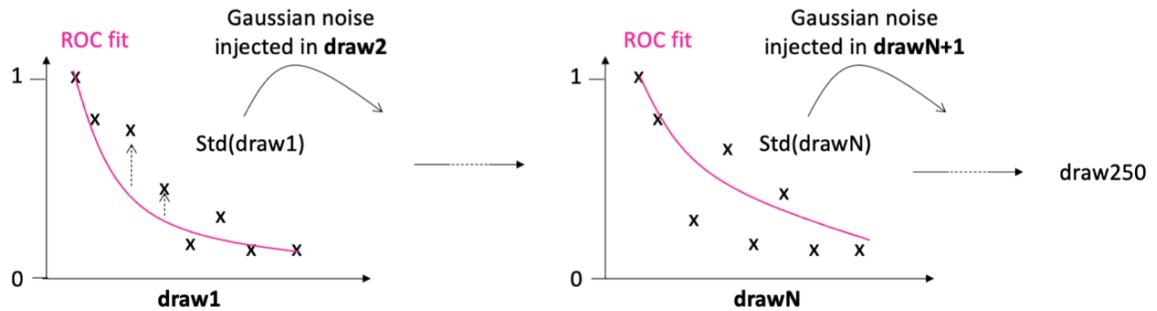

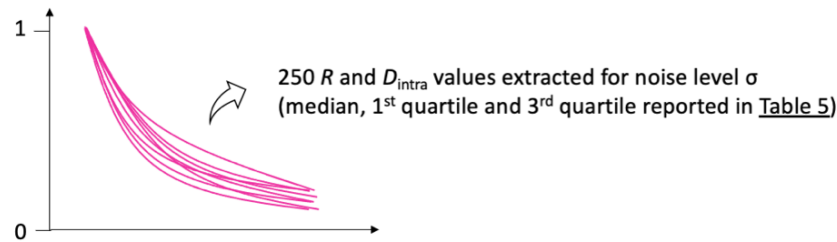

250 $R$ and $D_{intra}$ values extracted for noise level σ
(median, 1st quartile and 3rd quartile reported in Table 5)